\documentclass[pre,twocolumn,twoside,byrevtex,superscriptaddress,floatfix]{revtex4-2}

\usepackage{currfile}
\usepackage{graphicx,epsfig,verbatim,enumerate}
\usepackage{amssymb,amsmath}
\usepackage{ifthen}

\usepackage{longtable}

\usepackage{mathtools}
\usepackage{rotating}

\usepackage{array}

\newboolean{twocolswitch}
\usepackage{tcolorbox}
\tcbuselibrary{breakable}
\tcbuselibrary{most}
\newtcolorbox{dialogue}{
  colback=gray!5,
  colframe=gray!50,
  boxrule=0.5pt,
  arc=3pt,
  left=6pt,
  right=6pt,
  top=6pt,
  bottom=6pt
}

\newcommand{\sindex}[1]{}
\newcommand{\nindex}[1]{}

\newcommand{\www}[1]{\url{#1}}

\usepackage{lettrine}

\usepackage{color}

\setboolean{twocolswitch}{true}

\begin{document}

\title{\protect
Narrative Structure in Tropes: 
A Computational Analysis of `Friends'
}


\author{
\firstname{Shun}
\surname{Zhang}
}

\email{shun.zhang@uvm.edu}

\affiliation{
  Computational Story Lab,
  Vermont Complex Systems Institute,
  MassMutual Center of Excellence for Complex Systems and Data Science,
  Vermont Advanced Computing Center,
  University of Vermont,
  Burlington, VT 05405, USA.
  }

\author{
  \firstname{Tabia Tanzin}
  \surname{Prama}
}


\affiliation{
  Computational Story Lab,
  Vermont Complex Systems Institute,
  MassMutual Center of Excellence for Complex Systems and Data Science,
  Vermont Advanced Computing Center,
  University of Vermont,
  Burlington, VT 05405, USA.
  }

\author{
  \firstname{Christopher M.}
  \surname{Danforth}
}

\email{chris.danforth@uvm.edu}

\affiliation{
  Computational Story Lab,
  Vermont Complex Systems Institute,
  MassMutual Center of Excellence for Complex Systems and Data Science,
  Vermont Advanced Computing Center,
  University of Vermont,
  Burlington, VT 05405, USA.
  }

\affiliation{
  Department of Mathematics \& Statistics,
  University of Vermont,
  Burlington, VT 05405, USA.
  }

  \author{
  \firstname{Peter Sheridan}
  \surname{Dodds}
}

\email{peter.dodds@uvm.edu}

\affiliation{
  Computational Story Lab,
  Vermont Complex Systems Institute,
  MassMutual Center of Excellence for Complex Systems and Data Science,
  Vermont Advanced Computing Center,
  University of Vermont,
  Burlington, VT 05405, USA.
  }

\affiliation{
  Department of Computer Science,
  University of Vermont,
  Burlington, VT 05405, USA.
}
 
\affiliation{
  Santa Fe Institute,
  1399 Hyde Park Rd,
  Santa Fe, NM 87501, USA.
}

\affiliation{
  Complexity Science Hub,
  Metternichgasse 8,
  1030 Vienna,
  Austria
}

\date{\today}

\begin{abstract}
  \protect 

Tropes are recurring narrative devices in television and film. 
We carry out a computational analysis of tropes in the sitcom Friends, using human-curated trope annotations from `TVTropes', episode transcripts, and IMDb ratings. Because automatic trope detection remains challenging, we treat existing trope annotations as a curated analytical layer and focus on their downstream narrative and semantic functions. We first examine the relationship between episode-level trope frequency and audience reception. 
We find a statistically significant positive association between trope count and weighted IMDb ratings, although the modest explanatory power suggests that more than just trope density explains audience evaluation. We then connect trope annotations to dialogue transcripts and represent trope-related dialogue using TF-IDF-based semantic features. Using PCA and k-means clustering, we group 1,954 distinct tropes into 15 semantically interpretable clusters. Chi-square analyses show that the six main characters are unevenly distributed across these clusters, with character-specific trope profiles that are broadly consistent with their established narrative identities. Finally, we project trope clusters into the ousiometric power–danger space to examine their semantic organization. The results show that `Physical and Sexual Comedy' occupies a region associated with relatively high danger, while `Revelation, Surprise, and Reaction' occupies a region associated with relatively high power. 
Overall, our work first demonstrates a way to operationalize trope measurement, and second, shows that identifiable trope clusters can provide holistic `distant-reading' descriptions of characters and stories.

\end{abstract}

\maketitle

\section{Introduction}
Recurring narrative patterns constitute a fundamental component of storytelling of all kinds.
Originally framed as motifs for folktales,
such conventions for television and film 
came to be commonly referred to as tropes~\cite{GarcaSnchez2021TheSD, tvtropes_def}.
Tropes may be classic plots, short sequences of actions, jokes, or types of character interactions. Tropes are important because human beings view the world through structured narratives and accumulated experience~\cite{Bruner1991TheNC}. These narrative structures recur across a vast number of works, thereby shaping viewers’ expectations regarding 
character and plot development~\cite{Mittell2015ComplexTT}. 

Previous research has demonstrated that large collections of narratives consistently exhibit recurring structural patterns~\cite{Luomala1949TheHW, Reagan2016TheEA}. Many stories rely heavily on such repeated narrative frameworks~\cite{Propp1959MorphologyOT}, and tropes are among them. This leads us to our aim of quantitatively examining the structural functions of tropes in storytelling. Compared with themes or broad genres, tropes are more concrete but still more general than individual plot events. This intermediate level of abstraction makes them particularly suitable for our systematic analysis. 

Here, we use the “Brick Joke” trope as an example to illustrate the function of tropes as recognizable and repeatable narrative devices. The `Brick Joke' is a seemingly irrelevant feature set up early, and the punchline unexpectedly returns much later for comedic effect~\cite{tvtropes_brickjoke}. Some unrelated jokes can be told in between to add bonus points. When that feature was a joke, it was also known as `Chekhov's Gag' or `Call back'. It is widely used in media works and can be adapted to different contexts.

Recent computational studies have begun to investigate automatic trope detection and large-scale trope analysis~\cite{GarcaOrtega2020TropesIF, Flaccavento2025AutomatedDO}. Nevertheless, identifying tropes still remains a challenging task. Recent models perform poorly in detecting the tropes directly from content~\cite{Su2024InvestigatingVR}.

We adopt an alternative approach. We leverage existing human-curated trope annotations as a stable analytical ground truth and concentrate on downstream analysis. Specifically, we use the `TVTropes' dataset as our primary data source. `TVTropes' (tvtropes.org) is a large collaboratively curated online wiki that systematically documents recurring narrative tropes across television and film. The tropes data collected on this platform are all findings and summaries from real community members. Thus, its trope annotations provide a reliable foundation for examining how such tropes are linguistically realized through character dialogue and interaction.

Because tropes are highly context-dependent, variations in genre and production format can substantially influence their narrative function. To maintain stylistic consistency, we restrict our analysis to a single television series. We select `Friends' as our case study. Created by David Crane and Marta Kauffman, `Friends' is an American sitcom that aired from 1994 to 2004 and follows the lives of six friends in New York City across ten seasons and 236 episodes~\cite{friends_wikipedia}. As one of the most internationally recognized and popular television series (one of the highest-rated and most widely viewed television sitcoms on IMDb~\cite{imdb_friends_episodes}), `Friends' offers an especially suitable corpus for analysis, given its repeated use of recognizable narrative patterns. Therefore, we aim to investigate how tropes function to connect plot structures and character interactions in this show, which leads to the following research questions.

\begin{itemize}
    \item \textbf{RQ1: How is trope density associated with the episode ratings of Friends?}

Because tropes play a central role in narrative construction, it is natural to ask how strongly they matter to the show. Our first question is to investigate the relationship between trope density and episode quality. We use the episode as the unit of analysis. The number of tropes appearing in each episode is paired with its corresponding IMDb rating. In line with prior work that treats crowd-sourced ratings as a proxy for cultural reception~\cite{Ramos2015StatisticalPI, Wasserman2013CorrelationsBU}, the IMDb score can serve as an indicator of perceived episode quality, while the number of raters captures the reliability and reach of that evaluation. We use these measures to quantify episode popularity and to examine its relationship with trope frequency. 

    \item \textbf{RQ2: Do trope clusters correspond to character personalities?}
Although tropes are often treated as structural narrative devices, they are ultimately realized through characters and their interactions. Prior scholarship on television narratives has shown that long-running series frequently establish stable character identities that become recognizable to audiences over time~\cite{Mittell2015ComplexTT}. In Friends, the six main characters exhibit distinctive and widely recognized personality traits. We therefore investigate whether these established character identities are reflected in recurring trope patterns. 

Specifically, we explore whether characters exhibit distinctive distributions across trope clusters and whether those distributions align with established character identities. If trope participation systematically differs across characters, this would suggest that tropes contribute not only to plot organization but also to the expression and reinforcement of character identity. To answer this question, we need to do some data processing, since the raw dataset on tropes contains only labels and explanations. We need to connect them to the original transcripts and then explore trope clusters (details in the Data Section). 

    \item \textbf{RQ3: How do trope clusters behave in the ousiometric semantic space?}
Finally, we extend our analysis into an interpretable semantic framework, the ousiometric space~\cite{Dodds2026OusiometricsTE}. If tropes contribute meaningfully to characterization, their influence may also be visible at the level of language itself. To capture these underlying semantic patterns, we ask how tropes are represented in the ousiometric space. Specifically, we investigate how narrative patterns are located in the space and shift over time and across different story arcs and characters.
\end{itemize}

\section*{Data}
In this section, we explain the data and the preparation work for our research questions. For RQ1, we require episode-level trope counts along with IMDb ratings to examine the relationship between trope density and audience evaluation. For RQ2 and RQ3, we additionally require the full episode transcripts. By aligning the transcripts with the tropes, we can identify the characters involved in each trope and analyze how different characters participate in distinct trope clusters.

\subsection{Data Collection}
\subsubsection{`Friends' Scripts}
Our dataset consists of full dialogue transcripts from all ten seasons of the television sitcom \textit{Friends}, which are publicly available online~\cite{corbari_friends_scripts}. The corpus covers all 236 episodes aired between 1994 and 2004 and contains approximately $10^{6}$ tokens in total, with an average of about 4,000 tokens per episode. Each episode is transcribed into text with speaker attributions. Specifically, in some seasons, the last two episodes are combined into a single file for plot coherence. They are all divided into the correct episode division according to IMDb~\cite{imdb_friends_episodes}. A special case in the dataset involves clip show episodes. Since these episodes largely recycle previously aired dialogue, the associated tropes are often inherited from earlier contexts~\cite{Lotz2007TheTW}. To avoid duplication and bias in ratings, we identify and remove all clip show episodes for Research Question 1. Those include S4E21, S6E20, S7E21, S8E19. 

\subsubsection{`Friends' Tropes}
The trope dataset is collected from the website `tvtropes.org'~\cite{tvtropes_def}, a community-curated repository of recurring narrative patterns in film and television. For each Friends episode, community members manually document the tropes that appear, based on close reading of the episode’s narrative and dialogue. Each identified trope is accompanied by a short explanation describing how it is used in the show. The trope lists are organized alphabetically within each episode (236 episodes). The overall size of the tropes is 5,794.

\subsubsection{`Friends' Ratings}
For each episode, the IMDb ratings are collected together with the total number of user votes for each episode of `Friends'~\cite{imdb_friends_episodes}. Each episode has a rating from 1 to 10, accurate to one decimal place. The number of raters for each episode is collected to remove rating bias, with two significant figures. All clip show episodes have been removed from the dataset.

\subsection{Data Pre-processing}
The transcripts need to be cleaned and standardized before analysis. Scripts from all ten seasons of Friends are converted into a consistent plain-text format. Non-lexical content, such as scene markers (`[Scene: The Subway, Phoebe is singing for change.]') and stage directions (`(Ross gestures his consent.)'), is removed from the transcripts. Speaker normalization: Nicknames and abbreviated forms (e.g., `Rach' to `Rachel', `Pheebs' to `Phoebe') are standardized. These preprocessing steps produce a more consistent corpus for later analysis.

\subsection{Data Annotation}
The original dataset only provides episode-level trope annotations and explanations. We need to identify specific dialogue lines that correspond to each trope. The process is as follows. 

\subsubsection{Mapping scripts to tropes}
For each episode, all tropes documented in the `TVTropes' list are treated as candidates within each episode. Each trope has a paragraph of explanation. We employ large language models (LLMs) to annotate tropes. Their role is to map trope explanations to concrete dialogues within scripts, following explicit and conservative annotation rules (see Appendix~\ref{sec:Prompt} for the detailed prompting strategy used for the mapping). 

To choose a better model, we conduct a validation comparison to assess the reliability of trope annotation across three LLMs. Specifically, we select a subset of episodes and perform parallel trope-to-dialogue annotation using GPT-4-class, Gemini, and LLaMA under identical prompts and annotation constraints. We then evaluate model outputs through manual inspection against trope definitions by two independent annotators, including one linguist and one domain expert familiar with TVTropes, assessing accuracy by whether annotated dialogue lines constitute clear, accurate instances of the target tropes. Each trope-instance is scored as: Correct: The annotated dialogue line(s) clearly and unambiguously instantiate the trope definition, with sufficient contextual evidence. Incorrect: The dialogue is tangential, vague, or entirely unrelated to the trope definition. 

Table~\ref{tab:model_comparison} shows that GPT-4-class achieved substantially higher annotation accuracy than the other models, particularly in excluding vague, incidental, or contextually insufficient dialogue.
    \begin{table}[t]
    \centering
    \caption{Small-scale internal validation of trope annotation reliability across large language models}
    \label{tab:model_comparison}
    \begin{tabular}{lccc}
    \hline
    \textbf{Metric} & \textbf{GPT-4-class} & \textbf{Gemini} & \textbf{LLaMA} \\
    \hline
    Sampled Tropes & 50 & 50 & 50 \\
    Annotated Dialogue lines & 50 & 50 & 50 \\
    Correct Annotations & 45 & 35 & 32 \\
    Accuracy (\%) & \textbf{90.0} & 70.0 & 64.0 \\
    \hline
    \end{tabular}
    \end{table}

The annotation process is generally reliable, but several segmentation and parsing errors are identified. These issues are primarily caused by truncated strings, corrupted character encoding, and parsing failures during the alignment of trope annotations with episode transcripts. In some cases, quotation marks contained within trope descriptions are incorrectly interpreted as dialogue boundaries, resulting in misaligned segments. To minimize the impact, we manually inspected the data and corrected obvious mismatches where necessary. In total, 112 annotation records were manually repaired before the final dataset was constructed.

\begin{figure*}
  \centering
  \includegraphics[width=\textwidth]{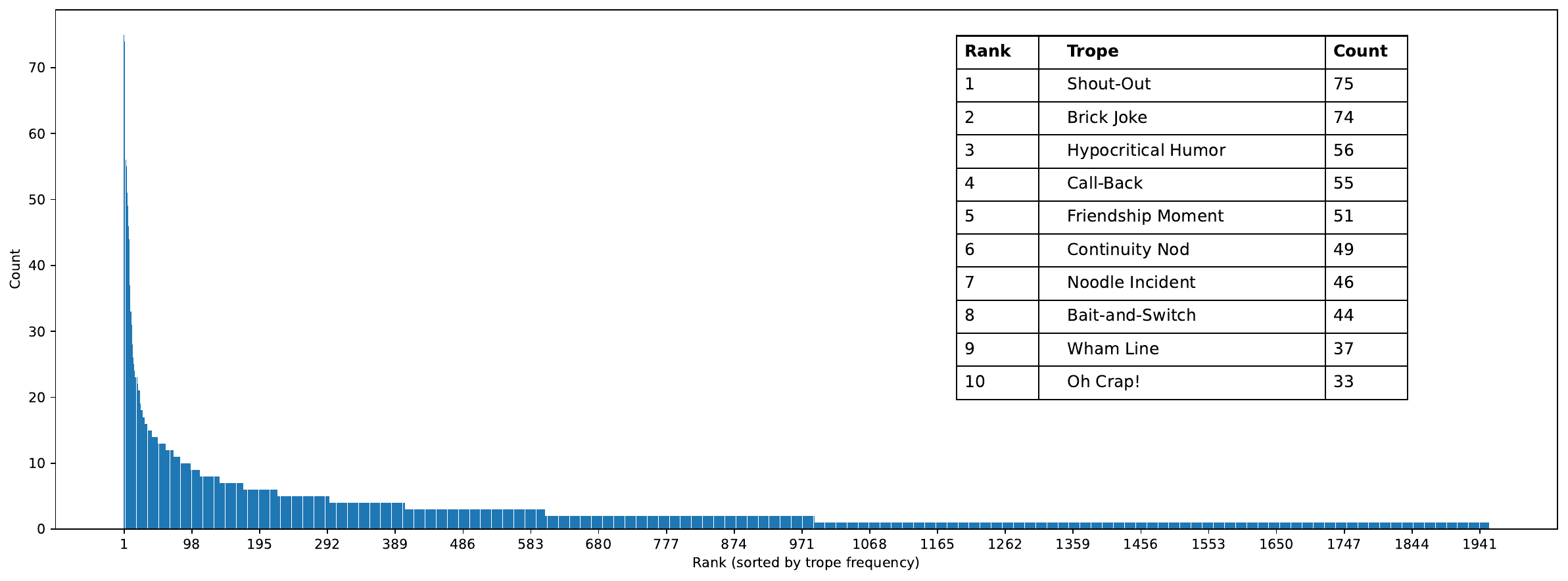}
  \caption{\textbf{Distribution of all tropes after annotation}. Inset: the ten tropes with the largest count. To be specific, a `Shout-Out' trope is an element in a piece of media that intentionally refers to something or someone outside the work, such as fans, family members of the cast or crew, or, commonly, another piece of pop culture~\cite{tvtropes_shoutout}.}
  \label{fig:trope_distribution}
\end{figure*}

After data cleaning and data annotation, the complete dataset contains 1,954 distinct tropes and 5,794 different dialogues. The overall distribution is shown in Figure~\ref{fig:trope_distribution}. 

\section{Methods}

\subsection{RQ1: How is trope density associated with the episode ratings of Friends?}
We want to investigate whether the density or configuration of tropes within episodes systematically relates to audience reception. We do so by exploring the relationship between the number of tropes in the show and its popularity, as measured by its ratings. We operationalize episode popularity using IMDb episode-level ratings. For each episode of Friends, we compute the total number of annotated tropes and their within-episode frequency distribution. Our analysis proceeds in two steps.

First, we conduct an exploratory visualization. We identify the highest-rated and lowest-rated episodes in the dataset and plot the distribution of tropes within each episode. These plots (Figure \ref{fig:tropesS01E20} and Figure \ref{fig:tropesS05E14}) show the frequency of tropes in the selected episodes, enabling a direct visual comparison of trope density between high- and low-rated episodes.

Then we employ a regression model to quantify the statistical association between trope density and episode ratings. The objective of this analysis is to examine whether trope-rich episodes tend to receive different audience evaluations rather than to establish a causal relationship between trope usage and episode popularity. 

To account for differences in rating reliability, we compute a weighted episode rating using an IMDb-style Bayesian average~\cite{Gelman2004BayesianDA}:
\begin{equation}
W_i = \frac{R_i v_i + C m}{v_i + m},
\end{equation}
where $W_i$ denotes the weighted rating of episode $i$, $R_i$ is the original IMDb rating, $v_i$ is the number of user ratings, $C$ is the mean rating across all episodes in the dataset, and $m$ is the median number of ratings. As the number of votes increases, the weighted rating converges toward the episode's original rating. Conversely, episodes with fewer votes are adjusted toward the overall average. To assess whether the observed association between trope frequency and episode ratings could be explained by broader seasonal differences, we additionally estimate a season fixed-effects model. Thus, we can write our regression model as:
\begin{equation}
W_i = \beta_0 + \beta_1 T_i + \sum_{s=2}^{10}\gamma_s D_{is} + \varepsilon_i,
\end{equation}
where $W_i$ is the weighted rating of episode $i$, $T_i$ is the number of tropes identified in that episode, $\beta_0$ is the intercept, $D_{is}$ is a dummy variable indicating whether episode $i$ belongs to season $s$. Season 1 serves as the reference category. The coefficient $\beta_1$ captures the association between trope frequency and episode ratings after accounting for season-level variation, while $\gamma_s$ represents the fixed effect associated with each season, and $\varepsilon_i$ is the error term. The fixed-effects specification therefore controls for unobserved season-level heterogeneity without imposing a linear assumption on seasonal differences.

\subsection{RQ2: Do trope clusters correspond to character personalities?}
Next, we want to examine whether these tropes also reveal deeper patterns in character construction. Specifically, if certain tropes recur around a character, they may reflect that character's underlying personality rather than serving as isolated narrative devices. 

To capture these patterns, we first construct TF-IDF-based vector embeddings of trope-related dialogue~\cite{Salton1988TermWeightingAI}. For each trope, all dialogue segments associated with that trope are converted into TF-IDF vectors to obtain a single vector representation of the trope. We employ TF-IDF representations because our objective is to identify recurring patterns across trope-related dialogue rather than to optimize semantic similarity performance. TF-IDF provides an interpretable representation of the vocabulary associated with each trope, making the resulting clusters easier to interpret qualitatively. Because TF-IDF representations are high-dimensional and sparse, we apply Principal Component Analysis (PCA) to reduce dimensionality before clustering. 

The resulting lower-dimensional (1000-dimensional) representations are then clustered using K-means clustering~\cite{Hartigan1979AKC}. Then we get a set of trope clusters that capture recurring semantic patterns. These trope clusters serve as the primary units of analysis in the subsequent character and semantic-space investigations.

To determine an appropriate number of trope clusters, we use four different methods to decide the cluster number, including the Elbow Method~\cite{Tibshirani2001EstimatingTN}, the Silhouette Score~\cite{Rousseeuw1987SilhouettesAG}, the Davies-Bouldin Index~\cite{Davies1979ACS}, and the Calinski-Harabasz Index~\cite{Calinski1974ADM}. Across these complementary criteria, a 15-cluster solution provided the most interpretable partition of the trope space (see Appendix~\ref{sec:optimal_cluster}, Figure~\ref{fig:optimal_cluster_number}, for the optimal cluster identification process).

Figure~\ref{fig:2D_clustering} and Figure~\ref{fig:2D_clustering_15} in the Appendix visualize the clustering results projected onto the first two dimensions. 

The cluster labels are assigned through manual inspection of representative tropes and trope descriptions within each cluster. Table~\ref{tab:trope_cluster_naming} shows the name of each trope cluster, the associated tropes, and the explanation of the naming.

\begin{table*}[t]
\centering
\small

\begin{tabular}{
>{\raggedright\arraybackslash}p{0.08\textwidth}
>{\raggedright\arraybackslash}p{0.2\textwidth}
>{\raggedright\arraybackslash}p{0.23\textwidth}
>{\raggedright\arraybackslash}p{0.40\textwidth}
}
\hline
\textbf{Cluster} 
& \textbf{Label} 
& \textbf{Typical Trope Representatives} 
& \textbf{Representative Trope Explanation} \\
\hline

0 & Casual Gag and Absurd Humor &
Brick Joke; Funny Background Event; Chekhov's Gag &
A Brick Joke is a joke that was set up early, and the punchline unexpectedly returns much later for comedic payoff. \\

1 & Strategic Conflict and Verbal Maneuvering &
Batman Gambit; Beat Them at Their Own Game; Blackmail &
Batman Gambit, a plan that revolves entirely around people doing exactly what you'd expect them to do. \\

2 & Emotional Escalation and Relational Crisis &
Belligerent Sexual Tension; Heroic BSoD; Gone Horribly Wrong &
Heroic BSoD: A stunning revelation or horrible event affects someone they care deeply about, leaving them shocked to the point of mentally shutting down for a while. \\

3 & General Sitcom Interactional Space &
A Family Affair; Accidental Declaration of Love; Act of True Love &
Accidental Declaration of Love. Just as it says literally, one unintentionally reveals romantic feelings during common situations. \\

4 & Authority, Dominance, and Social Regulation &
Death Glare; Fall Guy; Confronting Your Imposter &
The Death Glare is a "calm", murderous look often coupled with a tensed and menacing posture\\

5 & Cringe, Embarrassment, and Social Anxiety &
Bad Liar; Blind Date; Amazingly Embarrassing Parents &
Bad Liar is about a character who is unable to lie in a convincing way and comes up with really silly lies that fool no one. \\

6 & Character Development and Reflective Resolution &
Character Development; Bittersweet Ending; Friend-or-Idol Decision &
Character Development is just a character gradually changing in characterization or belief. \\

7 & Physical and Sexual Comedy &
Breast Attack; Bird-Poop Gag; Inconvenient Itch &
Bird-Poop Gag refers to the joke that a bird will poop on something or somebody inappropriately. \\

8 & Revelation, Surprise, and Reaction &
Didn't See That Coming; Delayed Reaction; Big ``WHAT?!'' &
Delayed Reaction occurs when a character reacts to something seconds later to cause a comedic effect. \\

9 & Competitive Escalation and Power Struggle &
Anything You Can Do; Escalating War; Cock Fight &
Escalating War refers to a rivalry where each side continuously intensifies to outdo the other. \\

10 & Romantic Negotiation and Identity Framing &
Be Yourself; Citizenship Marriage; Fairytale Wedding Dress &
Be Yourself is about someone accepting themselves for who they are \\

11 & Intimacy, Desire, and Moral Tension &
Commitment Issues; Everybody Has Lots of Sex; Fate Drives Us Together &
Commitment Issues: A character with commitment issues does not want to settle down with a Love Interest.  \\

12 & Family, Commitment, and Social Institutions &
Fallback Marriage Pact; Death by Childbirth; Brought Home the Wrong Kid &
Fallback Marriage Pact: Two unlucky-in-love Platonic Life-Partners or Friends with Benefits do not want to grow old alone and agree to marry each other if they're still single after a certain age.  \\

13 & Moral Resolution and Narrative Closure &
Both Sides Have a Point Remark; Happily Married; Ending by Ascending &
`Both Sides Have a Point' Remark is when two or more opposing sides of an argument eventually realize that each side has a point and that they shouldn't be fighting each other. \\

14 & Social Norm Violation and Self-Awareness &
Embarrassing First Name; Change the Uncomfortable Subject; I Need a Freaking Drink &
Change the Uncomfortable Subject is when some characters feel a current line of discussion is really inappropriate, and they make it clear they should move on to something else. \\

\hline
\end{tabular}
\caption{Semantic interpretation of trope clusters with representative evidence and trope explanations. All explanations are from `tvtropes.org'.}
\label{tab:trope_cluster_naming}
\end{table*}

To assess how characters are distributed across trope clusters, we also conduct chi-square tests comparing each character’s observed cluster distribution against the global expected distribution. The chi-square statistic is defined as
\[
\chi^2
=
\sum_{i=1}^{R}
\sum_{j=1}^{C}
\frac{(O_{ij}-E_{ij})^2}
     {E_{ij}},
\]
where $O_{ij}$ denotes the observed frequency of character $i$ in trope cluster $j$, and $E_{ij}$ denotes the expected frequency under the null hypothesis that character and trope-cluster are independent. 

To identify the character-cluster associations contributing most strongly to the overall chi-square statistic, we further compute and plot standardized Pearson residuals:
\[
r_{ij}
=
\frac{O_{ij}-E_{ij}}
     {\sqrt{E_{ij}}}.
\]
Positive residuals indicate that a character is associated with a trope cluster more frequently than expected under independence, whereas negative residuals indicate under-representation. We visualize it in a heatmap in Figure~\ref{fig:character_cluster_preference_heatmap}.

\subsection{RQ3: How do trope clusters behave in the ousiometric semantic space?}

Finally, we investigate how trope clusters behave within a semantic space. We select the ousiometric semantic space, a low-dimensional framework for semantic representation introduced by Dodds et al.~\cite{Dodds2026OusiometricsTE}. Ousiometry represents semantic meaning along three principal dimensions: Power, Danger, and Structure. Specifically, `Power' captures the degree of dominance and authority. The meaning runs from `void’ and `empty’ to `powerful’ and `success’. `Danger' captures risk and instability, in contrast to the direction of `safety' and `tranquility'. Structure is rather a less important dimension of the difference between `predictable' and `unpredictable'.

Alternative semantic frameworks, such as sentiment analysis or the Valence–Arousal–Dominance (VAD) model, primarily characterize emotional states~\cite{Warriner2013NormsOV}. However, trope clusters are narrative constructs rather than expressions of sentiment. Many tropes, including revelations, strategic conflicts, and social embarrassments, can occur in both positive and negative emotional contexts. Ousiometry is therefore particularly suitable because its dimensions emphasize power and danger, two properties central to narrative development rather than to emotional polarity alone.

Prior work has shown that ousiometric fluctuations can reveal structural transitions and semantic shifts within long-form narratives~\cite{Fudolig2022ADO}. More recent large-scale applications further demonstrate its ability to trace collective narrative dynamics and story-level semantic change across evolving discourse systems~\cite{Prama2025StoryAE}. In the context of Friends, power most likely reflects narrative agency and social control. Trope clusters involving authority, revelation, and conversational dominance tend to occupy higher-power regions. Danger is closer to social risk, which includes embarrassment and romantic conflict. Thus, those clusters associated with embarrassment or romantic uncertainty tend to occupy higher-danger regions. Although the structure dimension captures some valuable semantic organization, it is less directly related to the conflict-driven dynamics that motivate sitcom narratives. Narrative theory has long emphasized that stories are structured around changing relations of conflict and threat~\cite{Simmons2020RootNT}. This interpretation above aligns the Power–Danger framework with the core mechanics of sitcom storytelling. 

In our study, we therefore focus specifically on power and danger because they are most directly connected to narrative dynamics. Our objective is to examine how trope clusters are located in the Power–Danger space and how these semantic patterns vary across characters and seasons.

To project trope clusters into the Power-Danger space, we computed semantic scores for each dialogue line associated with a given trope. Specifically, for a dialogue segment \(d\) consisting of a sequence of words \(w_1, w_2, \ldots, w_n\), we calculate its Power score \(P(d)\) and Danger score \(D(d)\) as the mean of the Ousimetric score for all content words in the segment:

\begin{equation}
P(d) = \frac{1}{|W_d|} \sum_{w \in W_d} p_w, \qquad
D(d) = \frac{1}{|W_d|} \sum_{w \in W_d} d_w,
\label{eq:ousi_dialogue}
\end{equation}

where \(W_d\) denotes the set of content words present in the dialogue segment, \(p_w\) is the pre-computed Power score for word \(w\), and \(d_w\) is the pre-computed Danger score. For each trope cluster, we then aggregate across all constituent dialogue segments to obtain a cluster-level semantic centroid \((\bar{P}_c, \bar{D}_c)\):

\begin{equation}
\bar{P}_c = \frac{1}{|S_c|} \sum_{d \in S_c} P(d), \qquad
\bar{D}_c = \frac{1}{|S_c|} \sum_{d \in S_c} D(d),
\label{eq:ousi_cluster}
\end{equation}

where \(S_c\) denotes the set of all dialogue segments annotated as belonging to trope cluster \(c\). Season-level centroids are computed by subsetting \(S_c\) to dialogue segments from the corresponding season.

\section{Results and Discussion}
\subsection{RQ1: How is trope density associated with the episode ratings of Friends?}

\begin{figure}
      \centering
      \includegraphics[width=1.0\linewidth]{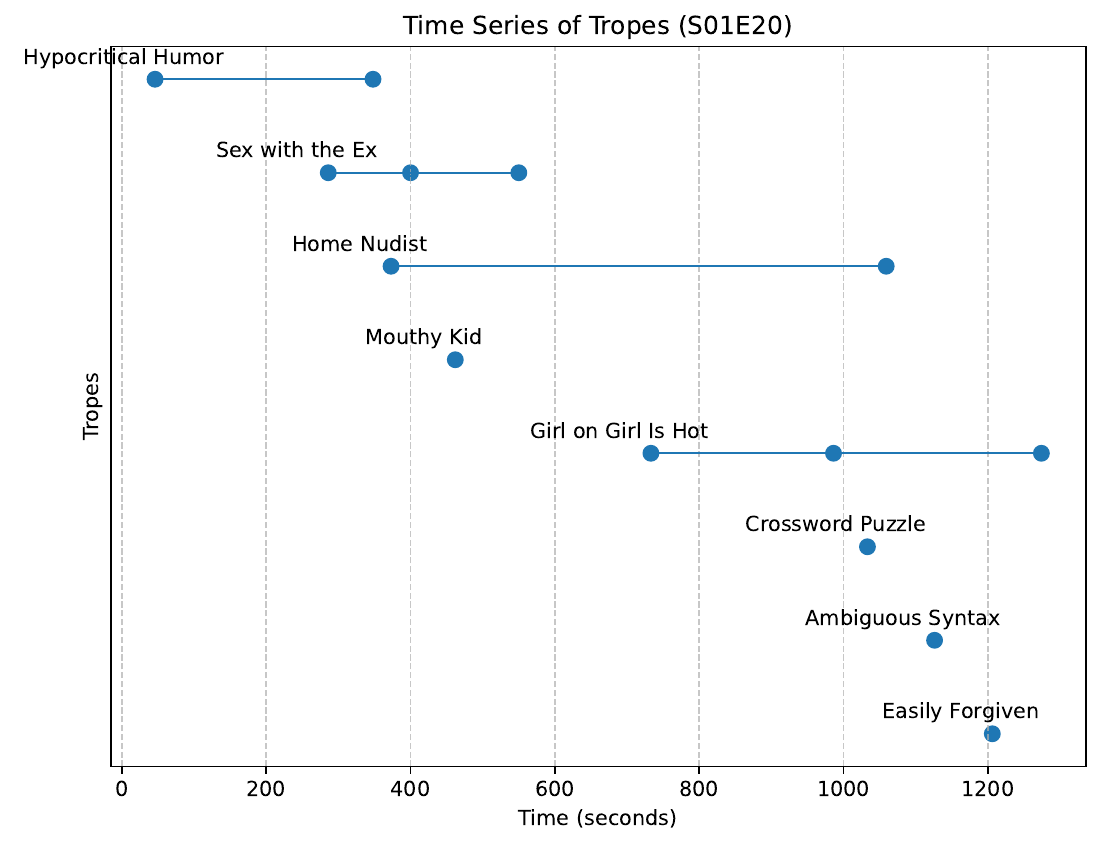}
      \caption{This figure visualizes the temporal distribution of narrative tropes across one of the lowest-rated episodes of Friends (S1E20). Each horizontal line corresponds to a distinct trope, with points indicating the timestamps (in seconds) at which the trope is instantiated in dialogue or scene context. When a trope occurs multiple times, the connecting line highlights its temporal span within the episode.}
      \label{fig:tropesS01E20}
    \end{figure}

\begin{figure}
      \centering
      \includegraphics[width=1.0\linewidth]{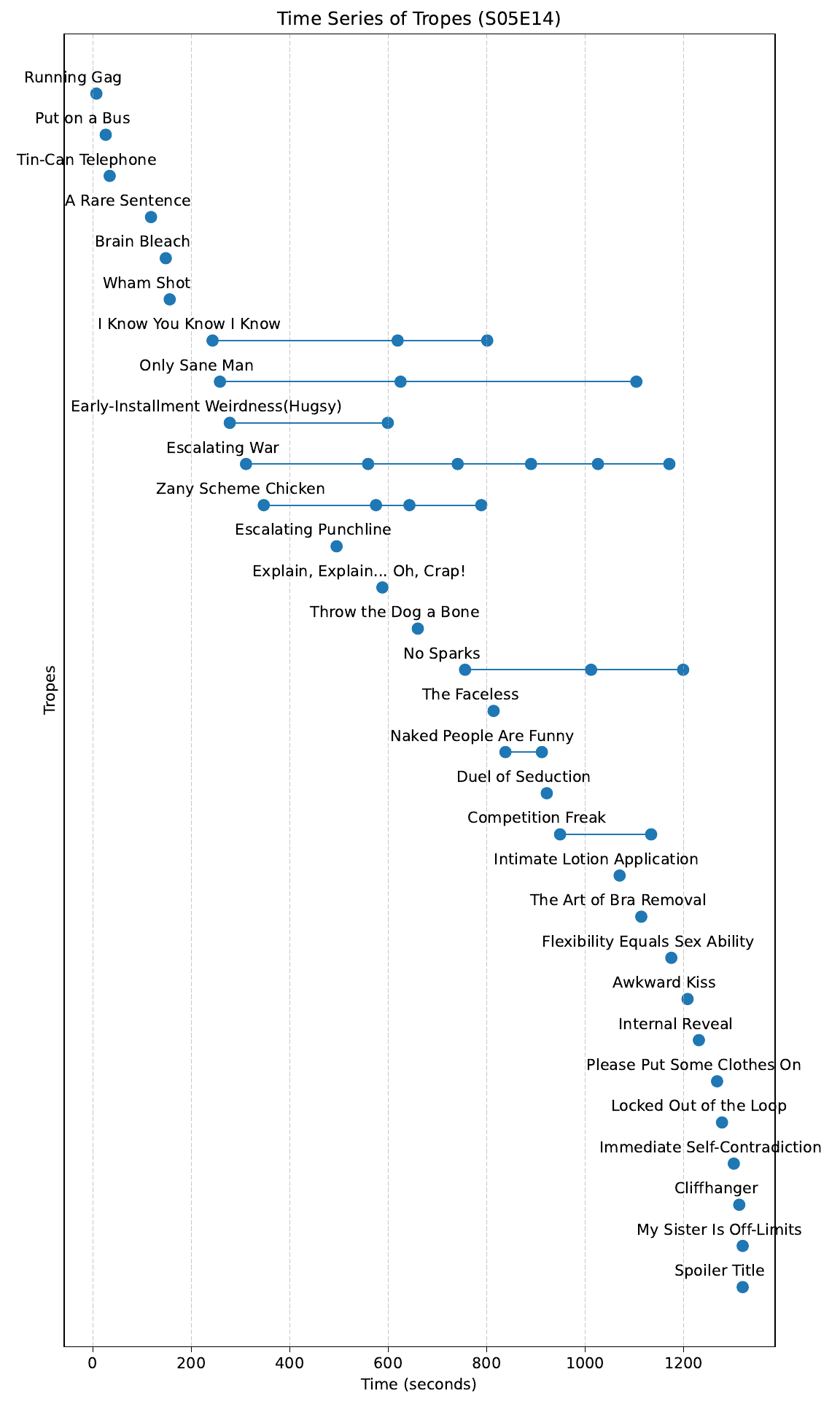}
      \caption{This figure visualizes the temporal distribution of narrative tropes across one of the highest-rated episodes of Friends (S5E14). Each horizontal line corresponds to a distinct trope, with points indicating the timestamps (in seconds) at which the trope is instantiated in dialogue or scene context. When a trope occurs multiple times, the connecting line highlights its temporal span within the episode.}
      \label{fig:tropesS05E14}
    \end{figure}

We select two representative episodes for qualitative comparison: one of the highest-rated episodes, Season 5 Episode 14 (IMDb rating: 9.7/10), and one of the lowest-rated episodes, Season 1 Episode 20 (IMDb rating: 7.7/10)~\cite{imdb_friends_episodes}. Figures \ref{fig:tropesS01E20} and \ref{fig:tropesS05E14} visualize trope occurrences as time series, showing when specific tropes occur throughout an episode. 

From Figures~\ref{fig:tropesS01E20} and \ref{fig:tropesS05E14}, we can observe that the higher-rated episode exhibits a greater density of trope occurrences throughout the narrative. In contrast, the lower-rated episode contains fewer distinct tropes. 
 
\begin{figure*}
      \centering
      \includegraphics[width=1.0\linewidth]{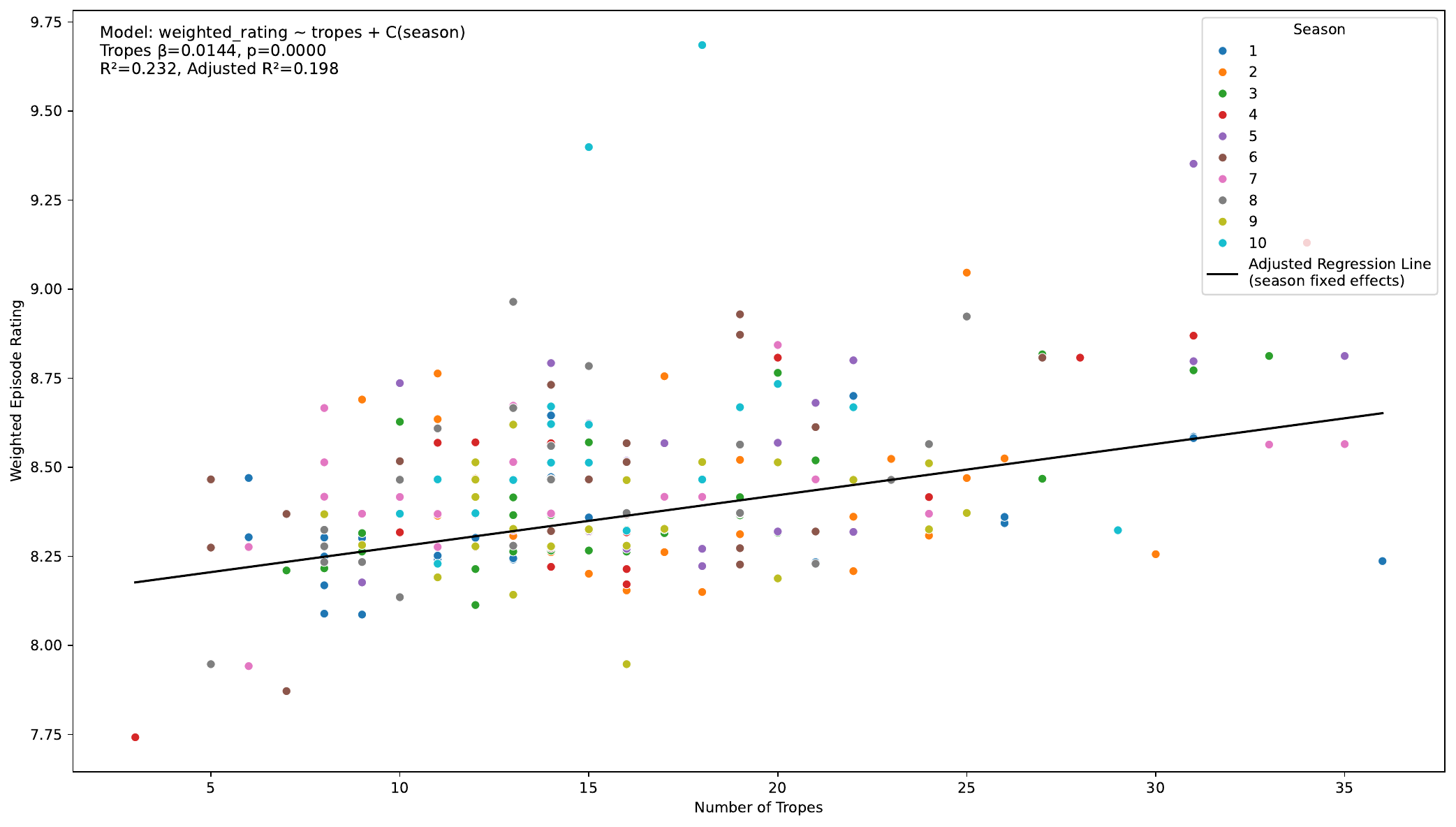}
      \caption{Each point represents one episode, plotted by the number of tropes and its weighted IMDb rating. Ratings are adjusted using a Bayesian weighted average that accounts for the number of raters. Colors indicate seasons. The black line represents the fitted relationship between trope frequency and weighted ratings from the season fixed-effects regression model. The model result is on the top left, indicating a significant positive association between trope frequency and ratings ($\beta_1=0.0144$, $p<0.0001$), with $R^2=0.232$ and adjusted $R^2=0.198$.}
      \label{fig:trope_number_by_rating_overall}
    \end{figure*}

Figure \ref{fig:trope_number_by_rating_overall} presents the results of the season fixed-effects regression relating trope frequency to weighted IMDb ratings. The fixed-effects specification controls for systematic differences across seasons.
The overall result (Figure \ref{fig:trope_number_by_rating_overall}) indicates a statistically significant positive association between trope density and episode ratings ($\beta_1 = 0.0144, p < 0.0001$). Episodes with more annotated tropes are associated with higher weighted IMDb ratings. Importantly, the positive association remains significant after controlling for season fixed effects, suggesting that the relationship between trope density and ratings cannot be explained only by differences among seasons. The model's explanatory power remains moderate ($R^2 = 0.232$), indicating that trope frequency represents only one component of audience evaluation. Episode ratings are likely influenced by additional narrative factors, such as story arcs or emotional payoff. This possibility is further explored through season-specific analyses. 

\begin{figure*}
      \centering
      \includegraphics[width=\textwidth]{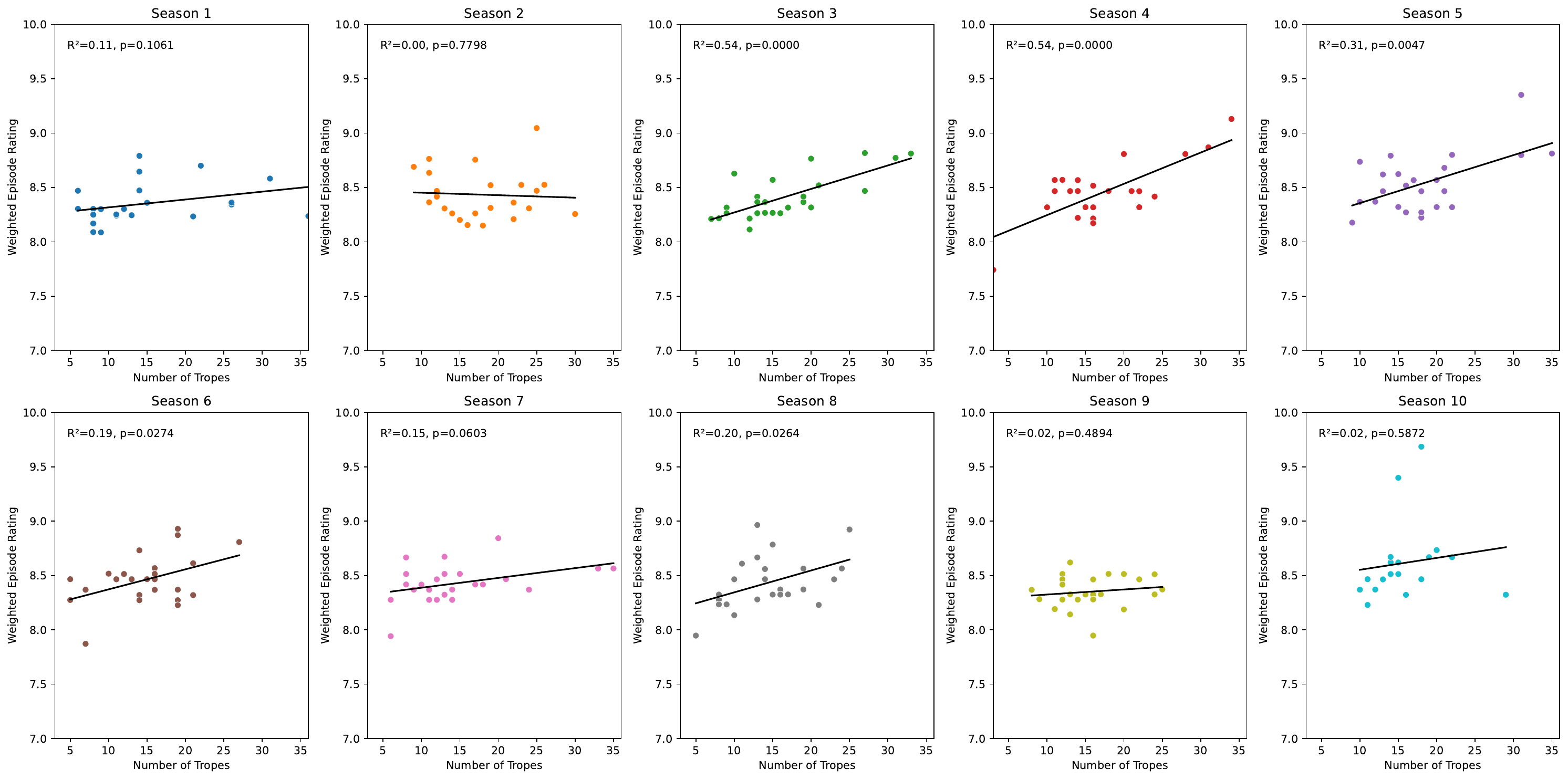}
      \caption{The result of the regression models for all ten seasons separately.}
      \label{fig:trope_number_by_rating_season}
    \end{figure*}

We further estimate season-specific regressions. Figure~\ref{fig:trope_number_by_rating_season} shows the result of the regression model applied to ten different seasons separately. 

Figure~\ref{fig:trope_number_by_rating_season} shows episode-wise scatter plots with fitted regression lines for each season. The strength of the association varies across seasons. Seasons 3 and 4 exhibit particularly strong relationships between trope frequency and weighted IMDb ratings ($R^2 = 0.54$ for both seasons), while Season 5 also shows a substantial association ($R^2 = 0.31$). In contrast, several other seasons display weaker or statistically nonsignificant relationships. 

This may suggest that the association between trope density and audience ratings is context-dependent. Identical trope counts may contribute differently to audience reception depending on how those tropes are integrated into character relationships and season-level narrative structures. 

One possible explanation is that the role of tropes changes throughout the series' lifecycle. In the early seasons, audience ratings may have been driven primarily by character establishment and relationship formation rather than by the density of recurring narrative devices. During Seasons 3–5, when the main characters and their relationships had already become established, recurring narrative devices like tropes may have played a larger role in shaping audience ratings. In later seasons, audience ratings may have become increasingly influenced by long-term story arcs, emotional resolution, and character closure, factors that are not fully captured by trope frequency alone.

\subsection{RQ2: Do trope clusters correspond to character personalities?}

\begin{figure*}
      \centering
      \includegraphics[width=\textwidth]{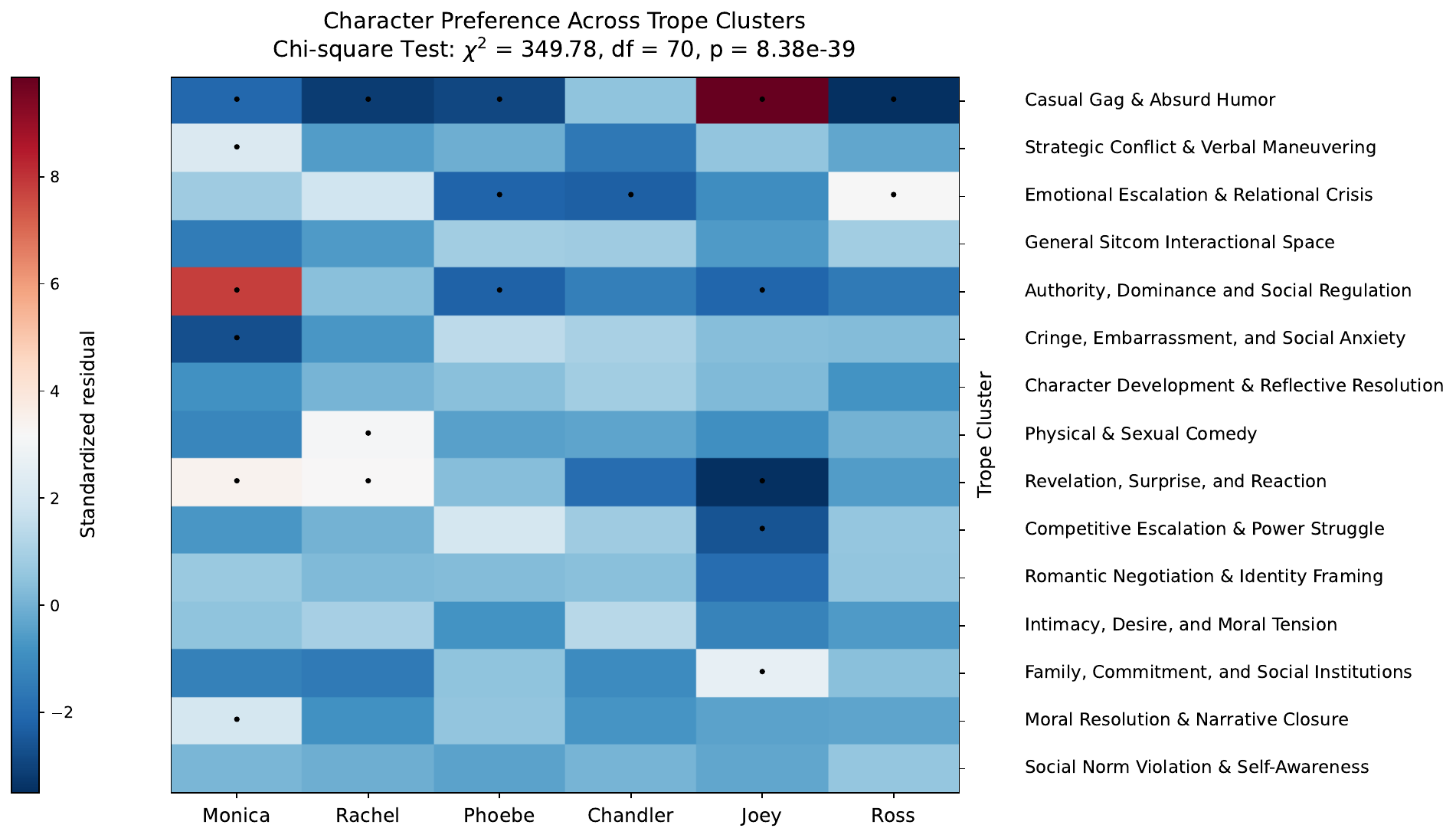}
      \caption{This heatmap shows standardized residuals from chi-square tests comparing each main character’s distribution across trope clusters to the global expected distribution. Positive values (red) indicate clusters in which a character appears more frequently than expected, while negative values (blue) indicate under-representation. Black dots mark cells with $\|\textbf{standardized residual}\| \geq 2$, highlighting statistically significant deviations.}
      \label{fig:character_cluster_preference_heatmap}
    \end{figure*}

We conduct a chi-square test comparing each character’s observed cluster frequencies with the global expected frequencies. We visualize the standardized residuals as a heat map, showing how the main characters (Monica, Rachel, Phoebe, Joey, Chandler, and Ross) differ in their participation across trope clusters in Figure~\ref{fig:character_cluster_preference_heatmap}. The test result is also shown in Figure~\ref{fig:character_cluster_preference_heatmap}. The large chi-square statistic $(\chi^{2} = 349.78, \text{df} = 70, p < .001)$ indicates that character participation across trope clusters differs substantially from what would be expected under a random distribution. Instead, each character exhibits distinctive trope preferences, suggesting that recurring narrative patterns are systematically associated with character identity and narrative role.

In addition, we have visualized the detailed distribution of the top 10 characters' participation across trope clusters in Figure \ref{fig:cluster_top10_characters_grid} (see Appendix).

Figure~\ref{fig:character_cluster_preference_heatmap} shows statistically significant deviations from expected trope distributions across the six main characters. Specifically, the black dots in the heat map mark character–cluster cells with absolute standardized chi-square residuals greater than 2. This distribution of the clusters for the six main characters aligns with their actual personalities in the show to a large extent.  

For example, Chandler occurs more often in the trope cluster \textit{Cringe, Embarrassment, and Social Anxiety}. This means Chandler participates more in this cluster, which reflects Chandler's recurring narrative role as a humor-oriented conversational deflector whose interactions are driven by sarcasm and verbal redirection. Social Anxiety and Embarrassment are often connected with sarcasm and awkwardness. Many scenes involving Chandler rely on self-deprecating humor or defensive joking in moments of emotional vulnerability. For example, there is an emotionally uncomfortable encounter with Monica’s ex-boyfriend Richard in S6E24, where Chandler gets awkward and anxious. This moment directly illustrates that Chandler's emotional vulnerability is repeatedly transformed into defensive humor.

\begin{quote}
\textbf{Richard:} \textit{Monica! Chandler!} \\
\textbf{Chandler:} \textit{Hey-hey, hey! I don’t know why I did that!} \\
\textbf{Monica:} \textit{Hey, it’s good to see you!} \\
\textbf{Richard:} \textit{You too, you let your hair grow long.} \\
\textbf{Monica:} \textit{Yeah—Oh that’s right. You always wanted me to. Hey, I see you got your mustache back.} \\
\textbf{Richard:} \textit{Well, my nose got lonely.} \\
\textbf{Chandler:} \textit{And uh, you don’t have a mustache, which is good.} \\
\textbf{Richard’s Date:} \textit{Hi, I’m Lisa.} \\
\textbf{Chandler:} \textit{Hi.} \\
\textbf{Richard:} \textit{Lisa, Monica, Chandler. We used to date.} \\
\textbf{Chandler:} \textit{Richard! No one’s supposed to know about us!} \\
\textbf{Richard:} \textit{...} \\
\textbf{Chandler:} \textit{See, I did it again.} \\
\textbf{Monica:} \textit{Chandler, why don’t we sit down?} \\
\textbf{Chandler:} \textit{Yeah, I’ll sit down.} \\
\textbf{Monica:} \textit{It’s good to see you.} \\
\textbf{Chandler:} \textit{I’m Chandler; I make jokes when I’m uncomfortable.} \\
\end{quote}

Also, when Chandler is actually about to propose, he becomes extremely nervous and emotionally overwhelmed, which leads to his absurdity. 

\begin{quote}
\textbf{Monica:} \textit{Are you okay?} \\
\textbf{Chandler:} \textit{Yes! I’m good! Are you good? Are you good? Is everything—are you—are you perrr-perfect?!} \\
\end{quote}
His reaction, with verbal derailment, aligns with the Verbal Maneuvering and Embarrassment/Social Anxiety clusters.

Next, Monica shows a strong association with the cluster \textit{Authority, Dominance and Social Regulation}. Throughout the series, Monica frequently initiates conflict by attempting to enforce standards or maintain social structure. For example, in S4E12, during the apartment trivia game, Monica tries to control the pace and remains highly competitive. She refuses to coordinate Rachel’s decisions and escalates the competitive stakes. Dialogues include
\begin{quote}
\textbf{Monica:} \textit{Rules are good! Rules help control the fun!} \\
\textbf{Rachel:} \textit{Monica, I don’t want to lose 200 dollars.} \\
\textbf{Monica:} \textit{We won’t.} \\
\textbf{Monica:} \textit{I own this game! Look at my hand.} \\
\textbf{Rachel:} \textit{Why? Do you have the answers written on there?} \\
\textbf{Monica:} \textit{No! Steady as a rock!} \\
\textbf{\ldots}\\
\textbf{Rachel:} \textit{I’m not moving! This is my apartment, and I like it!} \\
\textbf{Monica:} \textit{I’ll take care of it.}
\end{quote}
These moments illustrate that Monica's authority exhibits competitive dominance and control over interaction, which aligns with the trope clustering result for Authority and Dominance. 

Another significant result is Joey. Joey demonstrates a stronger concentration in \textit{Casual Gag \& Absurd Humor} than all the others, strongly matching his established characterization. Joey participates in a large proportion of tropes in cluster 0, Casual Gag. In the show, unlike Chandler's humor, Joey's humor is often driven by naive interpretation and immediate emotional response. For instance, Joey’s humor frequently emerges from naive interpretations of social situations. During a charity auction in S6E24, Joey mistakenly believes the bidding process is a guessing game.
\begin{quote}
\textbf{Mr.~Thompson:} \textit{The winning bid was a whopping \$20,000!} \\
\textbf{Joey:} \textit{I won! That was my guess!} \\
\textbf{Rachel:} \textit{Joey! It is an auction! You don’t guess, you buy!} \\
\textbf{Joey:} \textit{What?! I don’t have \$20,000!} \\
\textbf{Rachel:} \textit{What were you thinking?!} \\
\textbf{Joey:} \textit{I didn’t know it was an auction!} \\
\textbf{Rachel:} \textit{Why would a charity give away a free boat?!} \\
\textbf{Joey:} \textit{I don’t know! Charity?}
\end{quote}
These interactions derive humor from Joey’s inability to recognize institutional and conversational subtext. These absurd escalations and gag-oriented social habits are the reasons why Joey is strongly concentrated in the \textit{Casual Gag \& Absurd Humor} trope cluster.

Furthermore, these distributions suggest that those character personalities are clearly correlated with trope use based on our results. Different characters consistently occupy distinct regions of trope space, reflecting stable narrative roles.

\subsection{RQ3: How do trope clusters behave in the ousiometric semantic space?}

\begin{figure*}
      \centering
      \includegraphics[width=1.0\linewidth]{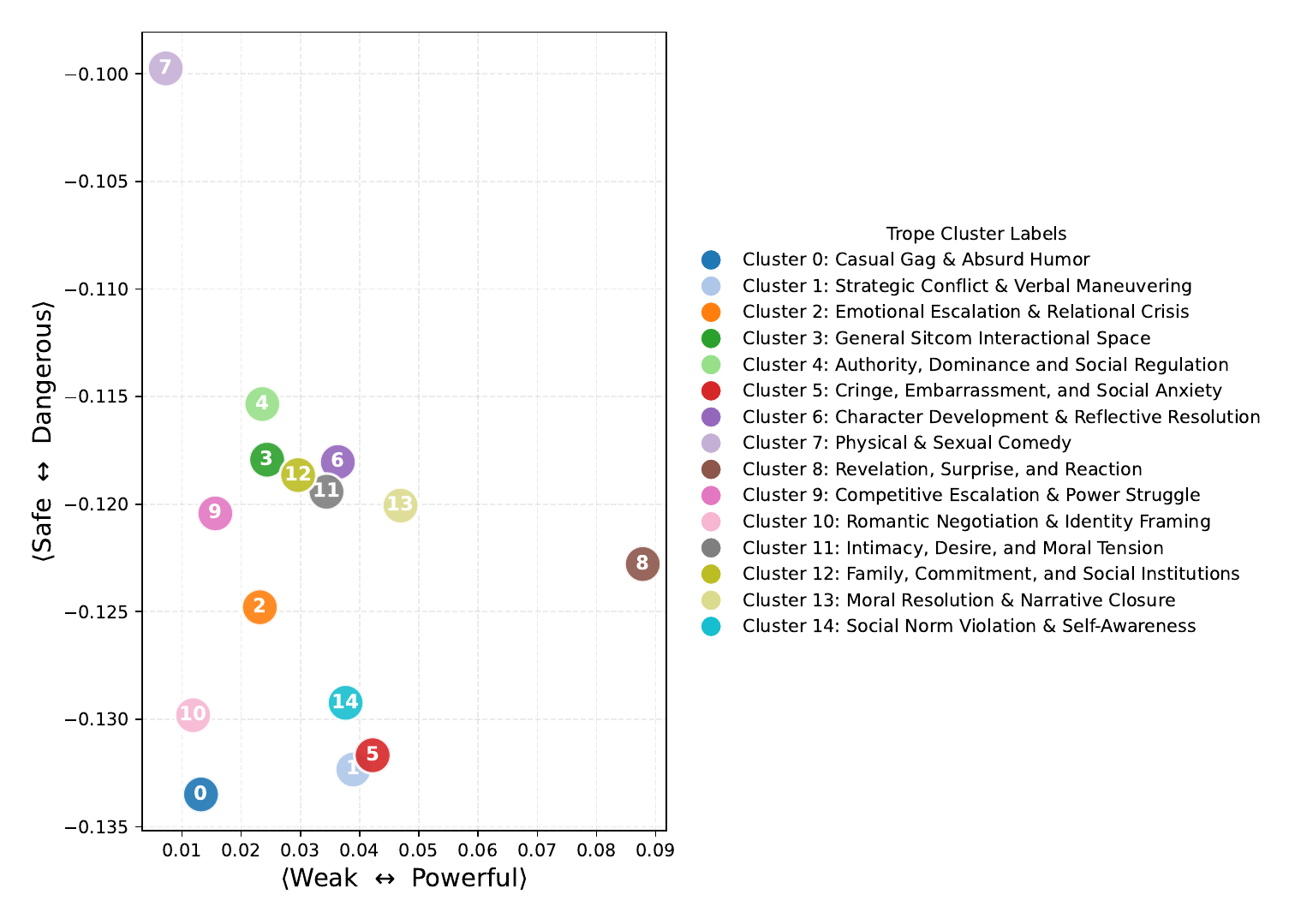}
      \caption{All-seasons Power--Danger distribution of trope clusters in \textit{Friends}. Each point represents the season-aggregated semantic centroid of a trope cluster in the ousiometric space, with the horizontal axis indicating relative power (weak $\leftrightarrow$ powerful) and the vertical axis indicating relative danger (safe $\leftrightarrow$ dangerous). The distribution reveals clear variation across trope clusters. We can notice that clusters 7 and 8 are distinct from others. Clusters 3, 4, 6, 9, 11, 12, and 13 are in the mid area among all clusters.}
      \label{fig:All_Seasons_Power_Danger_Distribution_of_Trope_Clusters}
    \end{figure*}

We project trope clusters into the Power–Danger space using semantic scores aggregated across all ten seasons. Figure \ref{fig:All_Seasons_Power_Danger_Distribution_of_Trope_Clusters} presents the all-seasons distribution of trope clusters, where each point denotes the overall semantic centroid of a cluster. The distribution is not random. Instead, trope clusters occupy distinct regions of the Power–Danger space, suggesting that trope clusters are organized according to underlying semantic functions.

From Figure~\ref{fig:All_Seasons_Power_Danger_Distribution_of_Trope_Clusters}, we can notice that clusters 7 and 8 are clearly in different positions from the others. Cluster 7 (Physical and Sexual Comedy) occupies the highest-danger region of the semantic space. In ousiometric terms, such situations are associated with instability and vulnerability rather than physical harm. The cluster occupies a high-danger region because it systematically encodes forms of social risk that are central to sitcom humor. On the other hand, Cluster 8 (Revelation, Surprise, and Reaction) occupies the highest-power region. Tropes such as \textit{Didn't See That Coming} and \textit{Big “WHAT?!”} revolve around the disclosure of previously hidden information. These moments often alter character knowledge or reshape social relationships, placing the cluster in a semantically powerful region of the ousiometric space. Clusters 3 (General Sitcom Tropes) and 5 (Cringe, Embarrassment, and Social Anxiety) are common tropes that occur for a large proportion, so that the traces are more condensed.

    \begin{figure*}
      \centering
      \includegraphics[width=1.0\linewidth]{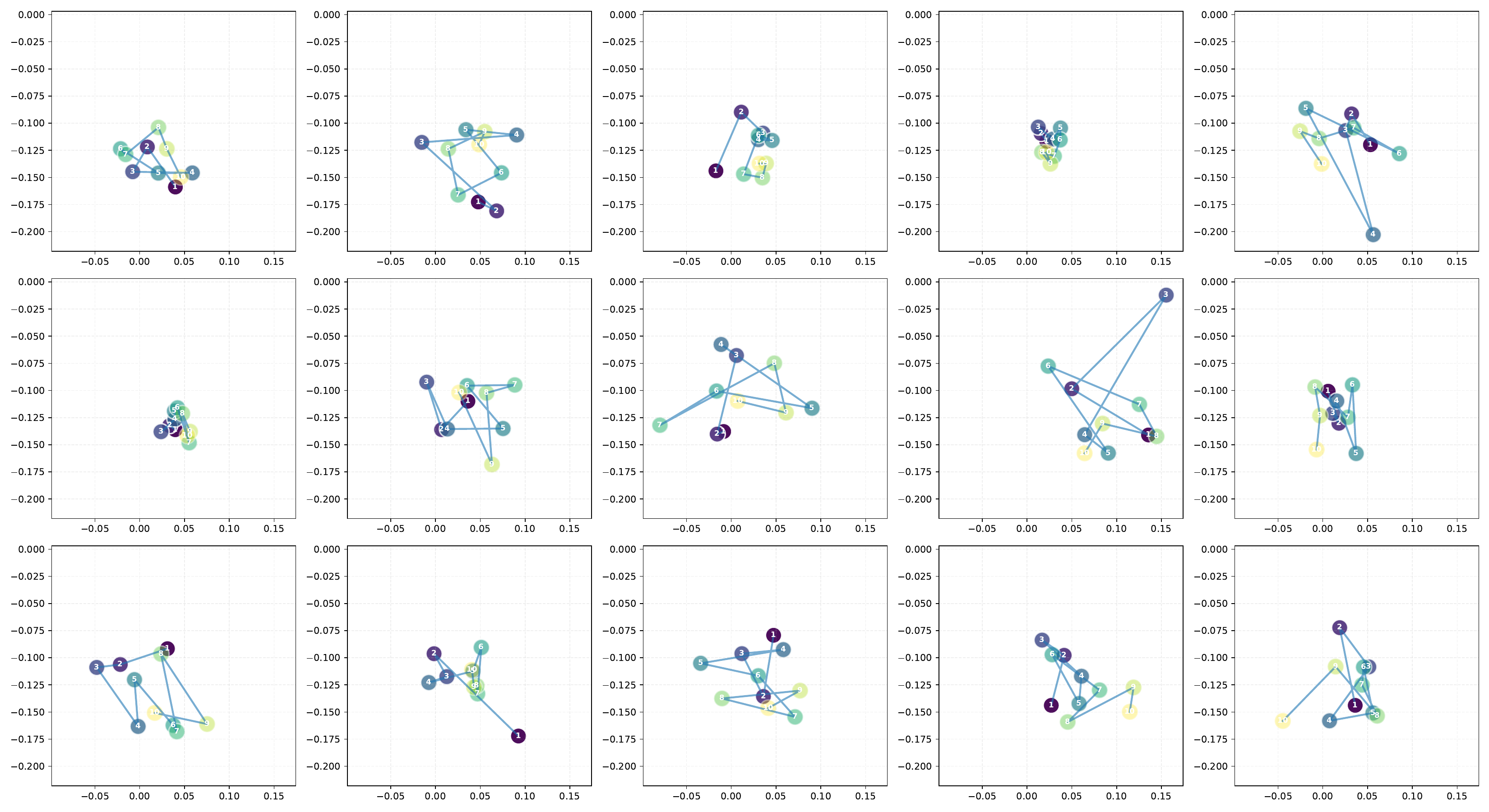}
      \caption{Power--Danger trajectories of trope clusters across all seasons of `Friends'. Each panel represents one trope cluster, showing its season-wise movement in the ousiometric semantic space, with power on the horizontal axis and danger on the vertical axis. Points represent seasonal centroids, and connecting lines trace temporal progression across seasons. The trajectories illustrate how recurring narrative patterns shift in semantic meaning over time.}
      \label{fig:Power_Danger_Trajectories}
    \end{figure*}

Figure \ref{fig:Power_Danger_Trajectories} further demonstrates that trope semantics are not static. Several clusters, particularly Clusters 4, 7, 8, and 10, exhibit substantial movement across seasons, indicating that similar tropes may be used in different semantic contexts over time. Specifically, the substantial movement of Cluster 7 (Physical and Sexual Comedy) may reflect the context-dependent nature of sitcom humor. In the early seasons, it often arises from awkward physical jokes, such as in S1E19, where Phoebe gets hit with the tranquilizer dart meant for Marcel, the monkey (Trope: Instant Sedation). 
\begin{quote}
\textbf{Phoebe}: \textit{Oh, this is so intense. One side of my butt is totally asleep, and the other side has no idea.}
\end{quote}
In Season 7, similar comedic devices increasingly occur within established romantic partnerships and family-oriented storylines. The danger score tends to become low, such as in S7E24, where Monica wants to have babies with Chandler right away (Trope label: My Biological Clock Is Ticking).
\begin{quote}
\textbf{Monica}: \textit{(Enters) Okay, it's baby time. Pants off, Bing. (Sees Ross) Didn't see you there, Geller.}\\
\textbf{Chandler}: \textit{Yeah, Ross is here so...}\\
\textbf{Ross}: \textit{Yeah, and I was really hoping that I could hang out...}
\end{quote}
Although the trope category remains the same, its semantic position within the Power–Danger space changes because the surrounding narrative context evolves.

\section{Conclusion}
This study investigates the structural role of tropes in the TV sitcom Friends through a computational analysis. Across all three research questions, the results demonstrate that tropes function as more than isolated storytelling devices. Instead, they operate as recurring narrative structures that connect audience engagement, character construction, and semantic organization.

For RQ1, we found a statistically significant positive association between trope density and episode ratings. Episodes with a higher density of annotated tropes tended to receive higher audience ratings. Although the observed relationship is robust, the model's modest explanatory power suggests that trope frequency is only one of many factors associated with audience evaluation. Also, our interpretation of the association between trope density and episode ratings (RQ1) is subject to potential reverse causality and annotator bias. While we observed a positive correlation, it remains plausible that higher-rated episodes attract more detailed curation from the TVTropes community, leading to a greater number of documented tropes rather than the tropes themselves driving the higher ratings. We therefore caution against interpreting the regression coefficient as a causal effect of tropes on audience reception. 

For RQ2, the clustering analysis revealed that trope participation is not distributed uniformly across characters. The six main characters exhibited distinctive associations with particular trope clusters, and these distributions closely aligned with their established narrative personalities. Recognizable character roles appear to be sustained through repeated participation in specific trope clusters. More importantly, the results indicate that character identity can be partially recovered from trope participation patterns alone. That is to say, personality information is possibly encoded in trope distributions. 

For RQ3, the ousiometric analysis revealed an additional semantic layer underlying the organization of tropes. Different trope clusters occupy distinct regions of the Power–Danger space, suggesting that recurring narrative patterns encode systematic semantic information rather than merely serving as structural labels. For example, `Physical and Sexual Comedy' is concentrated in high-danger regions, whereas `Revelation and Surprise' occupies high-power regions. Moreover, the seasonal trajectories of several clusters suggest that trope meanings are not entirely fixed but are shaped by evolving narrative contexts and character relationships.

Overall, these findings suggest that tropes function as an intermediate narrative unit linking local dialogue interactions, character personality construction, and higher-level semantic organization. Rather than descriptive labels, tropes capture recurring patterns of character personality constructions and information exchange that collectively shape narrative development. This study demonstrates that trope analysis provides a useful computational framework for understanding how stories generate meaning across multiple levels of narrative structure.

\section{Limitations and Future Work}

One limitation of this study is that only one television sitcom is examined. Some of the conclusions can be expanded to a broader scale. The correlation between tropes and ratings could be expanded to include all shows that have tropes collected on `tvtropes.org' and have IMDb ratings. Applying these methods to larger corpora would also allow direct investigation of the variability in trope interaction structures and further clarify how narrative roles emerge from character roles and from interactions between characters. 

Extending this framework beyond television to other narrative media, such as novels or theater plays, would further test the robustness of trope-level patterns. In that case, the trope types would probably be different from what we discuss in our study. Tropes in novels or plays might have fewer verbal and casual jokes, but more sophisticated plot constructions and subtle narrative devices. 

While we have shown that character roles remain relatively stable throughout the series, future research could model how character participation in trope clusters evolves across seasons or episodes, thus enabling a dynamic perspective on character development and transformation. For instance, we could examine whether shifts in trope participation precede changes in audience reception or narrative centrality. Future research could use time-based analyses to examine whether changes in trope configurations tend to come before shifts in story characteristics.

\acknowledgments

The authors acknowledge support from the National Science Foundation under Award No. 2242829.

\bibliographystyle{unsrt}
\bibliography{main_paper}

\section{Appendix}

\subsection{Prompt}
\label{sec:Prompt}

\begin{tcolorbox}[
  title=Trope Annotation Prompt,
  breakable,
  enhanced,
  colback=gray!5,
  colframe=black,
  listing only,
  listing options={
    basicstyle=\ttfamily\scriptsize,
    breaklines=true,
    breakatwhitespace=true,
    columns=fullflexible
  }
]
Instruction:
You are given:
1) A TV EPISODE SCRIPT\\
2) A LIST OF TV TROPES with DEFINITIONS\\

Your task is to identify dialogue lines that MATCH the trope definitions.\\

ANNOTATION RULES (STRICT)
- Select all dialogue lines that fit a trope definition.\\
- A trope may appear one or more times.\\
- A trope's occurrences may be non-sequential.\\
- All listed tropes are guaranteed to appear at least once.\\
- Do not force assignments.\\
- Quote dialogue exactly.\\

INPUT

Tropes:
${trope\_descriptions}$\\

Script:
${script}$\\

OUTPUT FORMAT (STRICT)
Return results in CSV-compatible format: ${Trope},{Dialogue}$\\

Each matching occurrence must be on a new line.\\

Example:
Cheer Up Episode,"It's okay, things will get better."
Brutal Honesty,"I don't care what you think."\\

IMPORTANT
- Output only the rows.\\
- Do not include headers.\\
- Do not add explanations.\\
- Do not add extra text.\\
\end{tcolorbox}

\subsection{Optimal Cluster Identification}
\label{sec:optimal_cluster}
\begin{figure*}
      \centering
      \includegraphics[width=\textwidth]{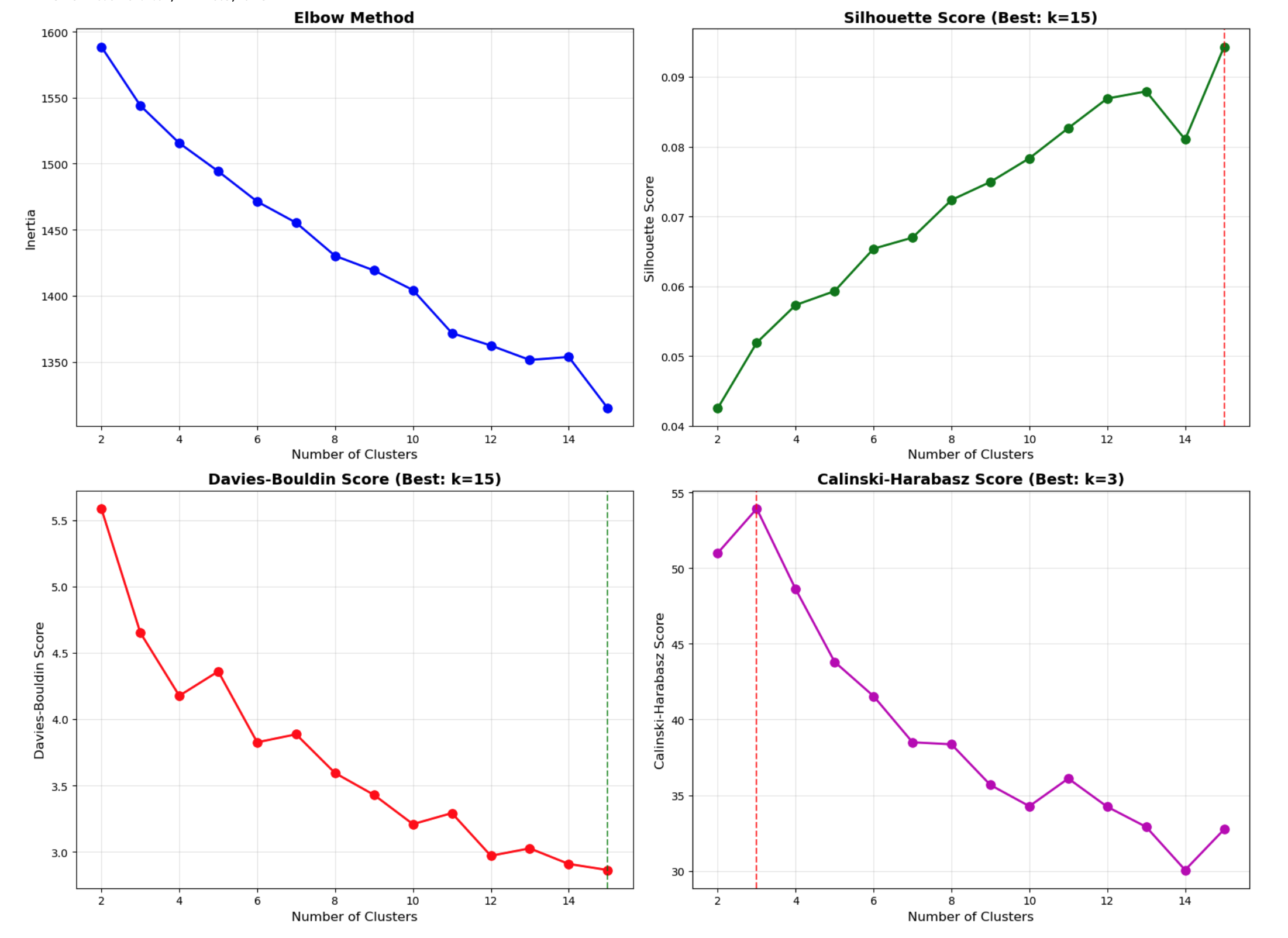}
      \caption{Evaluation of the number of trope clusters using four complementary criteria. The Elbow Method (top left) reports within-cluster inertia as a function of the number of clusters. The Silhouette Score (top right) and Davies-Bouldin Index (bottom left) both indicate optimal performance at $k=15$. The Calinski-Harabasz Index (bottom right) indicates smaller $k$ values.}
      \label{fig:optimal_cluster_number}
    \end{figure*}

As shown in Figure~\ref{fig:optimal_cluster_number}, both the Silhouette Score and the Davies-Bouldin Index indicate optimal performance at $k=15$. The Calinski-Harabasz Index indicates a smaller number. Considering our analytical goal of identifying interpretable yet sufficiently granular narrative units, we adopt a 15-cluster solution as a balanced choice.

Additional visualizations supporting these findings are provided in the Appendix. 

\subsection{Clustering Result Visualizations}
\begin{figure*}
      \centering
      \includegraphics[width=0.8\textwidth]{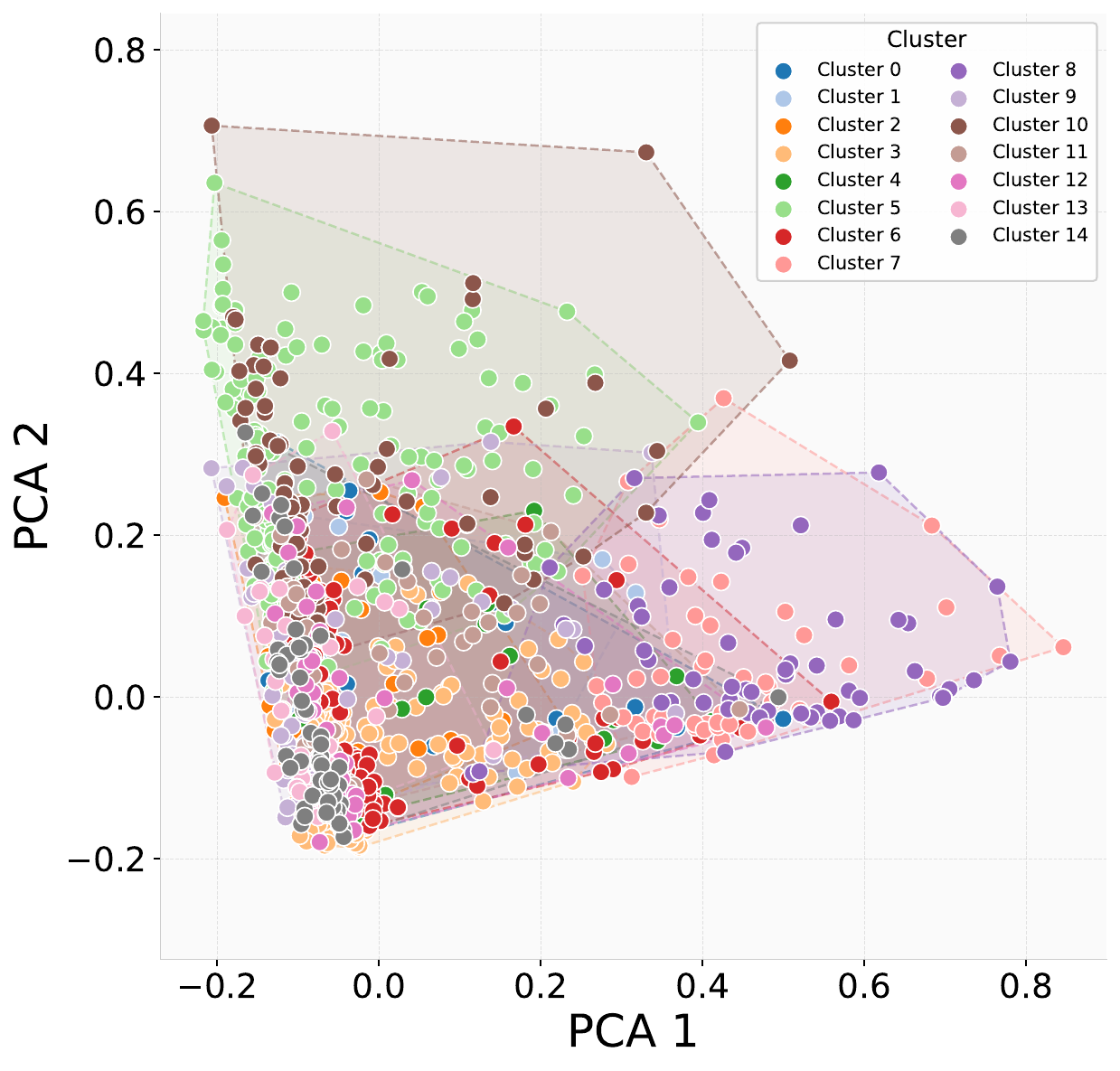}
      \caption{Distribution of trope clusters projected onto the first two principal components after PCA. Each point represents a trope, colored by its assigned cluster. }
      \label{fig:2D_clustering}
    \end{figure*}

Figure~\ref{fig:2D_clustering} shows the clustering results projected onto the first two dimensions. Considerable overlap is visible among clusters in the two-dimensional projection. This is expected because clustering was performed in the full 1000-dimensional PCA space, while the first two principal components explain only a small fraction of the total variance. The overlap suggests that trope clusters are not distinguished by a single dominant semantic axis. Instead, cluster structure emerges from combinations of many latent semantic dimensions. 
    
\begin{figure*}
      \centering
      \includegraphics[width=0.9\textwidth]{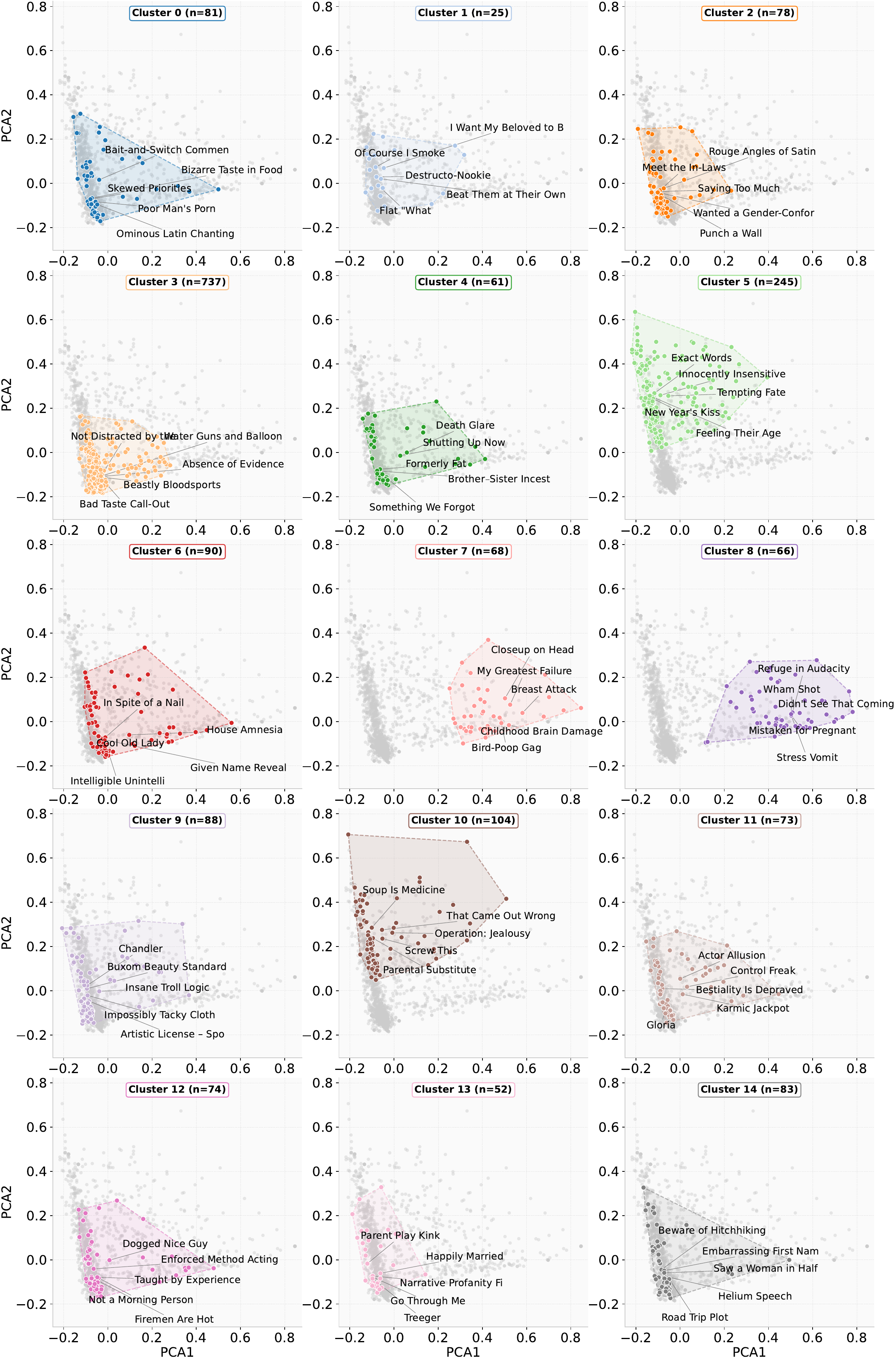}
      \caption{Clustering result: each cluster's position in the first two dimensions after PCA. }
      \label{fig:2D_clustering_15}
    \end{figure*}

Figure~\ref{fig:2D_clustering_15} shows the individual trope clusters projected onto the first two principal components. In each panel, colored points indicate tropes belonging to the cluster, while gray points represent all remaining tropes. The shaded polygon outlines the approximate extent of the cluster in the two-dimensional projection. Although clustering was performed in the full 1000-dimensional PCA space, the figure provides an intuitive view of the local semantic regions occupied by different trope clusters. Broad categories such as `General Sitcom Interactional Space' occupy a larger and more diffuse region, whereas more specialized clusters such as `Revelation, Surprise, and Reaction' are comparatively concentrated. 

\begin{figure*}
      \centering
      \includegraphics[width=\textwidth]{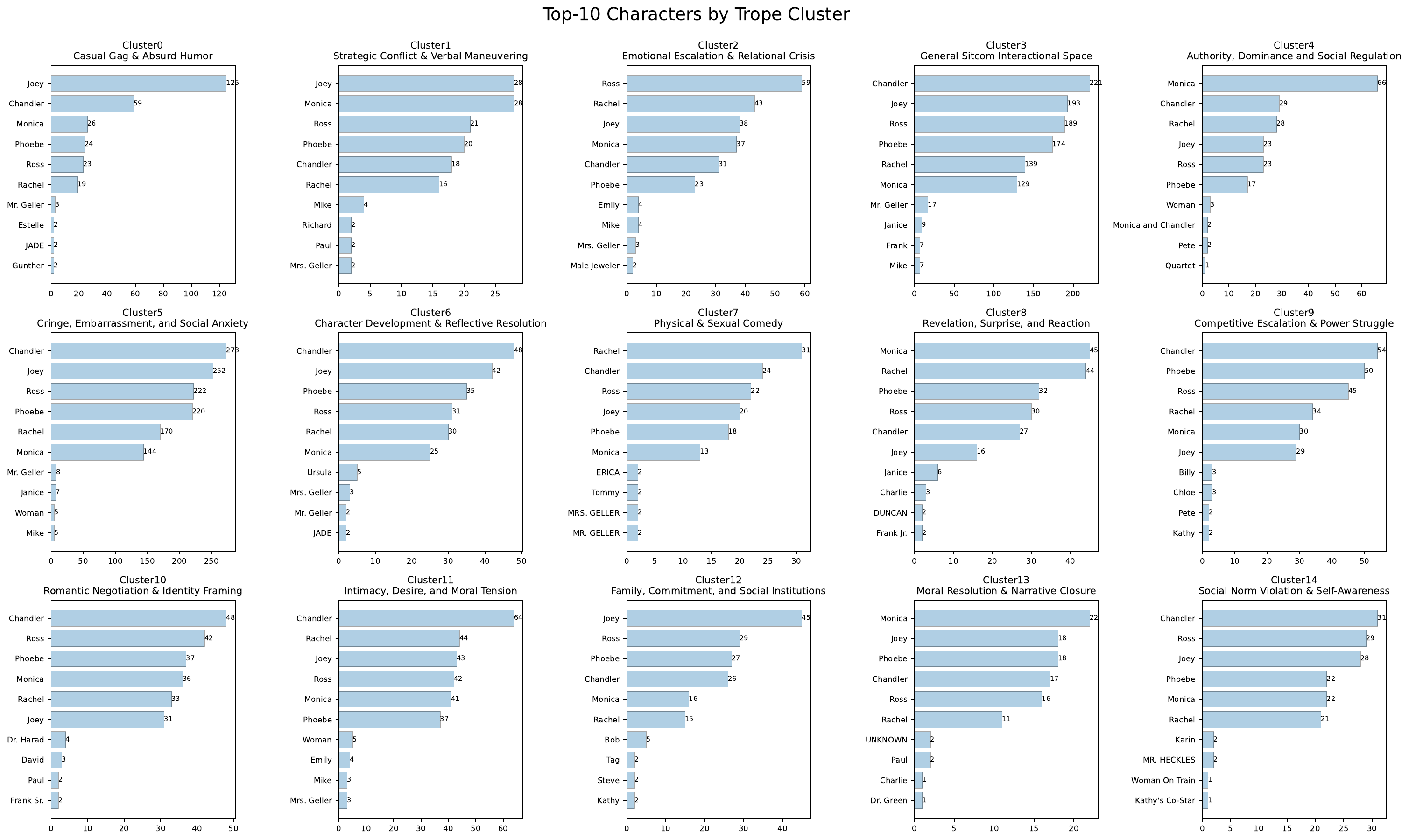}
      \caption{The top 10 characters' participation across trope clusters.}
      \label{fig:cluster_top10_characters_grid}
    \end{figure*}

Figure~\ref{fig:cluster_top10_characters_grid} visualizes the overall detailed distribution of the top 10 characters' participation across trope clusters.

\end{document}